\documentclass[fleqn,twoside]{article}
\usepackage{espcrc2}
\usepackage{graphicx}
\usepackage{color}

\newcommand{\bea}{\begin{eqnarray}}
\newcommand{\ena}{\end{eqnarray}}

\title{Bi-maximal mixing at GUT, the low energy data 
and the leptogenesis
\thanks{Presentation given by E. Takasugi}}

\author{
S. Kanemura\address[osaka]{Department of Physics, 
Osaka University Toyonaka, Osaka 560-0043, Japan},
 K. Matsuda\addressmark[osaka],
 H. Nakano\addressmark[osaka], 
 T. Ota\addressmark[osaka],
 T. Shindou\address{Theory group, KEK, Tsukuba 305-0801, Japan},
 E. Takasugi\addressmark[osaka]\thanks{takasugi@phys.sci.osaka-u.ac.jp}
 and K. Tsumura\addressmark[osaka]
}

\begin{document}

\begin{abstract}
In the framework of the minimum supersymmetric model with 
right-handed neutrinos, we consider 
the Bi-maximal mixing which is realized at the GUT scale and discuss a 
question that this model can reproduce 
the low energy phenomena and the leptogenesis. 
\end{abstract}
\maketitle
\vskip 5mm
\noindent
{\bf 1. The model} The neutrino masses are assumed to be derived 
through the seesaw mechanism, $m_\nu(M_X) = m_D^T D_R^{-1} m_D$, 
where $M_X$ is the GUT mass scale, 
$m_D$ is a Dirac mass matrix which is related to 
the neutrino-Yukawa coupling matrix $Y_\nu$ by 
$m_D = Y_\nu v\sin\beta/\sqrt{2}$. We assume that the 
right-handed neutrino mass matrix is a diagonal form 
as $D_R = \textrm{diag}(M_1,M_2,M_3)$. 

We assume that the neutrino mass matrix is diagonalized by 
the Bi-maximal mixing in the diagonal mass basis of charged 
leptons, i.e., $m_\nu(M_X)=O_BD_\nu O_B^T$, where $O_B$ is 
the Bi-maximal mixing matrix and 
$D_\nu= {\rm diag}(m_1,m_2e^{i\alpha_0}, m_3e^{i\beta_0})$ 
with $\alpha_0$ and $\beta_0$ being Majorana phases[1]. 

As for the neutrino mass spectrum, there are three cases, 
the hierarchical (H) masses, $m_1<<m_2<<m_3$, 
the inverse hierarchical (IH) masses, 
$m_1\simeq m_2>>m_3$ and the quasi-degenerate (QD) masses, 
$m_1\simeq m_2\sim m_3=m$. The renormalization group 
does not give any effect for the H case and this case is ruled out. 
Therefore, only the IH and QD cases may become consistent with the 
low energy data. 

The $\tau$-Yukawa coupling is known to contribute to rotate 
the solar angle from the Bi-maximal to the dark side[2]. 
Therefore, the other ingredient is required. 

We consider the effect of the neutrino-Yukawa coupling, 
$Y_\nu$ which enters 
in the renormalization group equation as $Y_\nu^\dag Y_\nu$. 
LFV processes, $l_j \rightarrow l_i + \gamma$, are 
induced by $|(Y_\nu^\dagger Y_\nu)_{ij}|^2$, so that 
its off-diagonal 
terms must be suppressed from the null observation. Therefore, 
we assume that $Y_\nu^\dag Y_\nu = {\rm diag} (y_1^2,y_2^2,y_3^2)$. 
It was found that the solar angle rotates to the normal 
side, only when $y_1^2-y_2^2>>|y_3^2-y_2^2|$[3][4]. This fact is shown 
later. With this assumption, 
$m_D$ is written by
\bea 
m_D=V_R^\dag D_DP_{ex}\;,\;
P_{ex} =\pmatrix{0&0&1\cr 0&1&0\cr -1&0&0\cr}\;,
\ena
where $D_D={\rm diag}(m_{D1},m_{D2},m_{D3})$ with 
$m_{D3}^2-m_{D2}^2>>|m_{D1}^2-m_{D2}^2|$ and 
$P_{ex}$ is a matrix to exchange the eigenvalues. 

In our previous paper[5], we considered the QD 
neutrino mass case, both for the hierarchical Dirac mass, 
$m_{D3}>>m_{D2}>>m_{D1}$. In this note, we consider both 
the IH and QD neutrino mass cases. 

By using the form of $m_D$, we find 
\bea
M_R^{-1} = D_D^{-1}
(P_{ex}O_B)D_\nu(P_{ex}O_B)^T D_D^{-1},
\ena
where $M_R\equiv V_R D_R V_R^T$. 
The diagonalization is easily made. 
We observe that $V_R$ and $M_i$ are determined
by three neutrino masses, two Majorana phases, 
and three Dirac masses $m_{Di}$, so that 
6 real positive masses and two Majorana phases. 
In addition to the masses, 
we use the data from the neutrino oscillations, 
which fix two neutrino masses. For simplicity, we 
assume the ratios of $m_{Di}$. Therefore, one mass scale 
$m_{D3}$ and two Majorana phases, and one mass scale for 
the QD case are free, so that all data can be computed 
with these parameters. 

\vskip 2mm
\noindent
{\bf 2. The input data} We take  $\tan^2 \theta_{@} \simeq 1$,
 $\tan^2 \theta_{\odot} \simeq 0.40$,
$\Delta m_@^2 =|m_3^2-m_2^2|\sim 2.5 \times 10^{-3}\ ({\rm eV}^2)$, 
$\Delta m_\odot^2=m_2^2-m_1^2 \sim 8.3 \times 10^{-5}\ ({\rm eV}^2)$. 
For the IH case, $m_3$ is not determined, and for the QD case, 
the overall mass is unknown. For the sake of argument, we take 
\bea
m_1&\simeq &m_2>> m_3 = 5.0\times10^{-3}{\rm eV}\;\;{\rm for \; IH}\;,
\nonumber\\
m_1&\simeq&m_2\sim m_3= 6.5\times10^{-2}{\rm eV}\;\;{\rm for \; QD}\;.
\ena 
For Dirac masses, we assume $m_{D1}/m_{D2}=m_{D2}/m_{D3}=1/5$ 
for simplicity.

Now the remaining unfixed parameters are $m_{D3}$ and 
two Majorana phases, aside from the parameters 
in the renormalization group, which 
we use $\tan \beta=20$, $M_X=2\times 10^{16}$GeV. 

For the solar neutrino angle, we find
\bea
\tan^2\theta_\odot =
\frac{1-2(2\epsilon_e-\epsilon_\tau)\cos^2(\alpha_0/2)m_1^2
/\Delta m_\odot^2}
{1+2(2\epsilon_e-\epsilon_\tau)\cos^2(\alpha_0/2)m_1^2
/\Delta m_\odot^2}\;,
\ena
where $\epsilon_e =(1/16\pi^2)(y_1^2-y_2^2)\ln(M_X/M_R)$, 
$\epsilon_\tau =(1/16\pi^2)((y_3^2-y_2^2)\ln(M_X/M_R)
 +y_\tau^2\ln(M_X/M_Z)$. 
From this, we find $2\epsilon_e-\epsilon_\tau>0$, which implies that 
$y_1^2>>y_2^2+y_3^2+y_\tau^2$, for this we introduced the matrix $P_{ex}$. 
In addition, we can derive $|\cos(\alpha_0/2)| \geq \cos2\theta_\odot
\sim  0.43$.  

\vskip 2mm
\noindent
{\bf 3. Predictions} \\
{\bf ${\bf M_i}$ and ${\bf m_{Di}}$:} The requirement that 
the low energy solar neutrino angle is reproduced 
determines the heavy neutrino mass 
$M_3$ and the Dirac mass $m_{D3}$ as a function of $\alpha_0$. 
The result is almost 
independent of $\alpha_0$ and $M_3\sim 1.4\times 10^{14}$GeV 
for the IH case, and $\sim 6\times 10^{13}$GeV for the QD case. 
The Dirac mass $m_{D3}$ is determined by 
$m_{D3}=\sqrt{m_2 M_3/|\cos(\alpha_0/2)|}$, which is of 
order 100GeV. 
The other Dirac masses are given by our assumption for 
Dirac masses. Also $M_1$ and $M_2$ are predicted once $\alpha_0$ is 
given. 

\vskip 2mm
\noindent
{\bf ${\bf |V_{13}|}$ and ${\bf \delta}$ :}
Ihe the Bi-maximal mixing, $V_{13}=0$ so that no $\delta$. In our case, 
these are induced by the renormalization group. $|V_{13}|$ is 
very small for the IH case and of order 0.01 for the QD case. 
$\delta+\beta_0$ is roughly between $-\pi/2$ and $-2\pi/3$.

\vskip 2mm
\noindent
{ ${\bf <m_\nu>}$:} 
The effective neutrino mass for the neutrinoless double beta 
decay is about $<m_\nu>\simeq m_2|\cos(\alpha_0/2)|
\ge 0.025$eV. 

\vskip 2mm
\noindent
{\bf Lepton Flavor Violation:} 
The LFV processes take place through the slepton mixing, 
which is proportional to $(Y_{\nu}^\dag \mathcal{L}Y_{\nu} )_{ij}$, 
where $\mathcal{L}={\rm diag}(\ln(M_X/M_1), \ln(M_X/M_2),\ln(M_X/M_3))$. 
Since $m_D^\dag m_D$ is diagonal, the contribution 
occurs only 
through $(Y_{\nu}^\dag (\mathcal{L}- <\mathcal{L}>)Y_{\nu} )_{ij}$. 

The branching ratios are given by the standard form and the result is 
$\textrm{Br}(\mu\rightarrow e\gamma)>>
 \textrm{Br}(\tau\rightarrow e\gamma)
>>\textrm{Br}(\tau\rightarrow \mu\gamma)$. 
We show the branching ratio for the IH case in Fig.1. 
The $\beta_0$ dependence of branching ratios comes in through 
the ratios of heavy neutrino masses. The branching ratios 
have the sizable dependence on Majorana phases, $\alpha_0$ 
and $\beta_0$.
\begin{figure}
\begin{center}
\includegraphics[width=4cm]{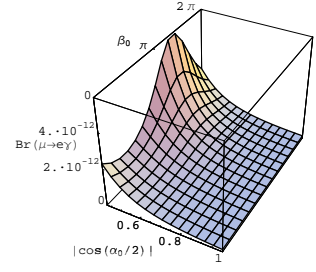}
\includegraphics[width=4cm]{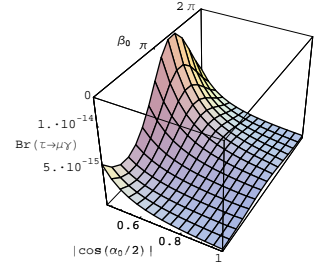}
\end{center}
\vskip-15mm
\caption{The branching ratio for the IH case 
 as functions of $|\cos(\alpha_0/2)|$ and $\beta_0$
 for $m_3=5.0\times10^{-3}{\rm eV}$. 
 We take $\tan\beta = 20$, $m_S = 300{\rm GeV}$,
 $m_0 =A_0=100{\rm GeV}$.}
\label{fig.1}
\end{figure}
 A similar behaviors are obtained for the QD case. 

\vskip 2mm
\noindent
{\bf Leptogenesis:} We have 
\bea
\epsilon_1 =-\frac{3m_3M_1}{4\pi v^2}
 \frac{\cos(\alpha_0/2)}{\sqrt{2(1+\cos\alpha_0)}  }
\left[\sin\left( \frac{\alpha_0}{2}-\beta_0 \right) \right]
\;,
\ena
for the IH case. For the QD case, we find
\bea
\epsilon_1 =
\frac{3\Delta m_{13}^2 M_1}{16 \pi v^2 m} \frac{\cos(\alpha_0/2)}{R}
\sin \left(\frac{\alpha_0}{2}-\beta_0 \right)\;,
\ena
where $\Delta m_{13}^2=m_1^2-m_3^2$ and 
\bea
R = \sqrt{1+\cos^2\frac{\alpha_0}{2}
+2\cos\frac{\alpha_0}{2}
\cos\left(\frac{\alpha_0}{2}-\beta_0\right)}.
\ena
For both cases, the asymmetric parameter depends on Majorana 
phases as $\cos(\alpha_0/2)\cos(\alpha_0/2-\beta_0)$ and 
small mass factor $m_3$ for the IH case, and 
$\Delta m_{13}^2=m_1^2-m_3^2$ for the QD case. 
The obtained values of $\eta_B$ is given in 
Fig.2 for IH and Fig.3 for QD. 
It is possible to reproduce the experimental result, 
$\eta_B \simeq 6\times 10^{-10}$. 
\begin{figure}
\begin{center}
\includegraphics[width=6cm]{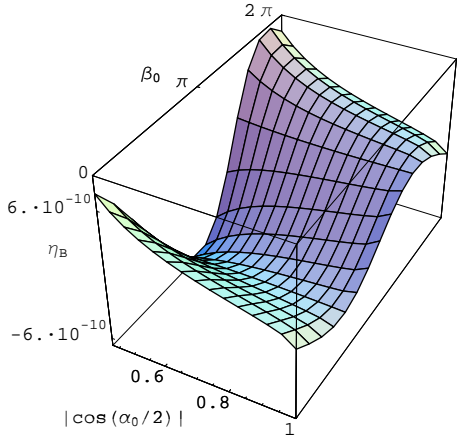}
\end{center}
\vskip-15mm
\caption{$\eta_{B}$
 as functions of $|\cos(\alpha_0/2)|$ and $\beta_0$
 for the IH case with $m_3 = 5.0\times10^{-3}{\rm eV}$.}
\label{Fig.2}
\end{figure}
\begin{figure}
\begin{center}
\includegraphics[width=6cm]{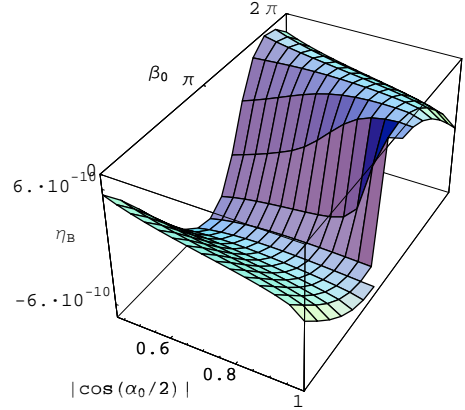}
\end{center}
\vskip-15mm
\caption{$\eta_{B}$
 as functions of $|\cos(\alpha_0/2)|$ and $\beta_0$
 for the QD case with $m = 6.5\times10^{-2}{\rm eV}$
 and $\Delta m_{13}^2>0$.}
\label{Fig.3}
\end{figure}
\vskip 2mm
\noindent
{\bf 4. Summary} We discuss the Bi-maximal mixing which is realized 
at the GUT scale. Our question is whether the Bi-maximal 
mixing can cope with low energy data and the leptogenesis. 
We showed that this is possible in the case where (11) element of 
$Y_\nu^\dag Y_\nu$ is much larger than other elements. 
From this and the LFV must be suppressed, we assume that 
$Y_\nu^\dag Y_\nu$ is diagonal. Then, CP phases including the 
Dirac phase are induced by two Majorana phases. This model 
enables to relate the CP phase which appears in the neutrino 
oscillation to the one in the neutrinoless 
double beta decay and also the CP phase in the leptogenesis.  
We showed the model is consistent with all experimental data. 





\begin{thebibliography}{99}
\bibitem{Maj}
S.~M.~Bilenky, J.~Hosek and S.~T.~Petcov,
Phys.\ Lett.\ B {\bf 94}, 495 (1980);
M.~Doi, T.~Kotani, H.~Nishiura, K.~Okuda and E.~Takasugi,
Phys.\ Lett.\ B {\bf 102}, 323 (1981);
J.~Schechter and J.~W.~Valle,
Phys.\ Rev.\ D {\bf 22}, 2227 (1980);
Phys.\ Rev.\ D {\bf 23}, 1666 (1981).
\bibitem{MST}T. Miura, T. Shindou, E. Takasugi, 
Phys. Rev. D66 (2002) 093002;
Nucl.Phys. A721, 537-540 (2003).
\bibitem{AKLR} S. Antusch, J. Kersten, M. Lindner and M. Ratz, 
Phys. Lett. B544 (2002), 1.
\bibitem{MST2} T. Miura, T. Shindou, E. Takasugi, 
Phys. Rev. D 68, 093009 (2003).
\bibitem{ST}T. Shindou, E. Takasugi, Phys.Rev. D70 (2004) 013005.
\end{thebibliography}
\end{document}